\documentclass[useAMS,usenatbib,usegraphicx]{mn2e}
\usepackage{epstopdf, txfonts, color, soul,hyperref}

\begin{document}

\title[The energy budget of stellar magnetic fields]{The energy budget of stellar magnetic fields}

\author[V. See et al.]
{V.See$^1$\thanks{E-mail: wcvs@st-andrews.ac.uk}, M. Jardine$^1$, A. A. Vidotto$^2$,  J.-F. Donati$^{3,4}$, C. P. Folsom$^{5,6}$,
\newauthor S. Boro Saikia$^{7}$, J. Bouvier$^{5,6}$, R. Fares$^{8}$, S. G. Gregory$^1$, G. Hussain$^{9}$, 
\newauthor S. V. Jeffers$^{7}$, S. C. Marsden$^{10}$, J. Morin$^{11}$, C. Moutou$^{12,13}$, 
\newauthor J. D. do Nascimento Jr$^{14,15}$, P. Petit$^{3,4}$, L. Ros\'{e}n$^{16}$, I. A. Waite$^{10}$\\
$^1$SUPA, School of Physics and Astronomy, University of St Andrews, North Haugh, KY16 9SS, St Andrews, UK\\
$^2$Observatoire de Gen\`eve, Universit\'e de Gen\`eve, Chemin des Maillettes 51, Sauverny, CH-1290, Switzerland\\
$^{3}$Universit\'{e} de Toulouse, UPS-OMP, Institut de Recherche en Astrophysique et Plan\'{e}tologie, Toulouse, France\\
$^{4}$LATT–UMR 5572, CNRS, Institut de Recherche en Astrophysique et Plan\'{e}tologie, 14 Avenue Edouard Belin, F-31400 Toulouse, France\\
$^{5}$Univ. Grenoble Alpes, IPAG, F-38000 Grenoble, France\\
$^{6}$CNRS, IPAG, F-38000 Grenoble, France \\
$^{7}$Universit\"at G\"ottingen, Institut f\"ur Astrophysik, Friedrich-Hund-Platz 1, 37077 G\"ottingen, Germany\\
$^8$INAF- Osservatorio Astrofisico di Catania, Via Santa Sofia, 78 , 95123 Catania, Italy\\
$^{9}$ESO, Karl-Schwarzschild-Str. 2, D-85748 Garching, Germany\\
$^{10}$Computational Engineering and Science Research Centre, University of Southern Queensland, Toowoomba, 4350, Australia\\
$^{11}$LUPM-UMR5299, CNRS and Universite ́ Montpellier II, Place E. Bataillon, Montpellier F-34095, France\\
$^{12}$Canada-France-Hawaii Telescope Corporation, CNRS, 65-1238 Mamalahoa Hwy, Kamuela HI 96743, USA\\
$^{13}$Aix Marseille Universit\'{e}, CNRS, LAM (Laboratoire d'Astrophysique de Marseille) UMR 7326, 13388, Marseille, France\\
$^{14}$Departmento de F\'isica Te\'orica e Experimental, Universidade Federal do Rio Grande do Norte, CEP:59072-970 Natal, RN, Brazil\\
$^{15}$Harvard-Smithsonian Center for Astrophysics, Cambridge, Massachusetts 02138, USA\\
$^{16}$Department of Physics and Astronomy, Uppsala University, Box 516, Uppsala 751 20, Sweden}

\maketitle

\begin{abstract}
Spectropolarimetric observations have been used to map stellar magnetic fields, many of which display strong bands of azimuthal fields that are toroidal. A number of explanations have been proposed to explain how such fields might be generated though none are definitive. In this paper, we examine the toroidal fields of a sample of 55 stars with magnetic maps, with masses in the range 0.1-1.5$\,{\rm M}_\odot$. We find that the energy contained in toroidal fields has a power law dependence on the energy contained in poloidal fields. However the power index is not constant across our sample, with stars less and more massive than 0.5$\,{\rm M}_\odot$ having power indices of 0.72$\pm$0.08 and 1.25$\pm$0.06 respectively. There is some evidence that these two power laws correspond to stars in the saturated and unsaturated regimes of the rotation-activity relation. Additionally, our sample shows that strong toroidal fields must be generated axisymmetrically. The latitudes at which these bands appear depend on the stellar rotation period with fast rotators displaying higher latitude bands than slow rotators. The results in this paper present new constraints for future dynamo studies.
\end{abstract}


\section{Introduction}
\label{sec:Intro}
Stellar magnetic fields are ubiquitous across the main sequence, playing a crucial role in the evolution of stars. Zeeman-Doppler imaging (ZDI), a tomographic imaging technique, has been used to map the large-scale surface fields of a wide variety of stars such as rapid rotators \citep{Donati1997,Donati1999AB,Donati1999HR,Donati2003Dynamo}, M dwarfs \citep{Donati2008a,Morin2008,Morin2010}, solar-type stars \citep[e.g.][]{Petit2008} and planet-hosting stars \citep[e.g.][]{Fares2013}. This technique is capable of identifying the field geometry as well as the energy budget of the different field components (axisymmetric/non-axisymmetric, dipole/quadrupole/higher order multipoles, poloidal/toroidal). Additionally, repeated observations of individual stars allows the temporal evolution of the field to be assessed with magnetic field polarity reversals observed for a number of stars \citep[e.g.][]{Donati2003,Donati2008b,Fares2009,Petit2009,Morgenthaler2011}. The wealth of data available has revealed how the magnetic properties of stars vary with various stellar parameters such as Rossby number and age \citep{Vidotto2014}.

Despite the progress being made in the field of stellar magnetism, there are still a number of open questions. For instance, what mechanisms are responsible for dynamo action and where does the dynamo operate? In the Sun, it is thought that fields are mainly generated in a thin layer between the radiative core and the convective zone, known as the tachocline, where shearing is strongest \citep{Charbonneau2010}. However, the same process cannot be responsible for field generation in all stars. Below the fully convective limit ($\sim$0.35$\,{\rm M}_\odot$), stars do not possess a tachocline and hence different dynamo mechanisms must be at work. Additionally, strong toroidal azimuthal fields have been observed on a range of stars \citep[e.g.][]{Donati1997,Donati1999HR,Petit2009,Morgenthaler2012,Jeffers2014,Saikia2015}. As highlighted by \citet{Jardine2013}, such azimuthal bands can only exist if the magnetic field has been stressed above its lowest energy state. Additionally, these authors determine that stellar winds cannot be the source of these stresses. Other authors have suggested that such azimuthal fields are evidence of dynamos distributed throughout the convection zone \citep{Donati1997,Donati2003,Donati2003Dynamo} rather than being confined to the tachocline as in the Sun. This view is supported by the work of \citet{Brown2010} who demonstrate that a tachocline is not required to generate strong bands of toroidal fields in rapidly rotating solar-like stars.

The toroidal component of stellar magnetic fields contains the free energy that, once liberated, is responsible for energetic events. Flares, coronal mass ejections and space weather in general have a large influence on the stellar environment, and can affect any planets orbiting the star \citep{Zarka2007,Griessmeier2007,Llama2011,Vidotto2012,Vidotto2013,Vidotto2015,See2014,See2015,Cohen2015}. Presently, the toroidal component of stellar magnetic fields has only been studied in single, or small samples, of stars. \citet{Petit2008} studied a sample of four solar-like stars and noted that the stellar rotation period plays an important role in determining the fraction of magnetic energy in the toroidal component of the stellar field. However, the rotation period cannot be the sole parameter that determines the toroidal energy fraction since stars with similar rotation periods show different toroidal energy fractions. Additionally, observations of individual stars, over multiple epochs, show that the toroidal energy fraction can change significantly on the time-scale of years \citep{Donati1997,Donati1999AB,Donati2003,Petit2009,Fares2010,Morgenthaler2012,Fares2013,Jeffers2014,Saikia2015}.  This indicates that the dynamo is dynamic in nature and cannot be characterised by single time averaged parameters. 

In this work, we conduct a statistical study of toroidal fields using a sample of 55 stars with ZDI maps. We discuss our sample of stars and the mathematical representation of their magnetic fields in section \ref{sec:Sample}. In section \ref{sec:Results} we cover the trends seen within our sample  with concluding remarks given in section \ref{sec:Discussion}.

\begin{figure}
	\begin{center}
	\includegraphics[trim=1cm 1cm 1cm 1cm, width=\columnwidth]{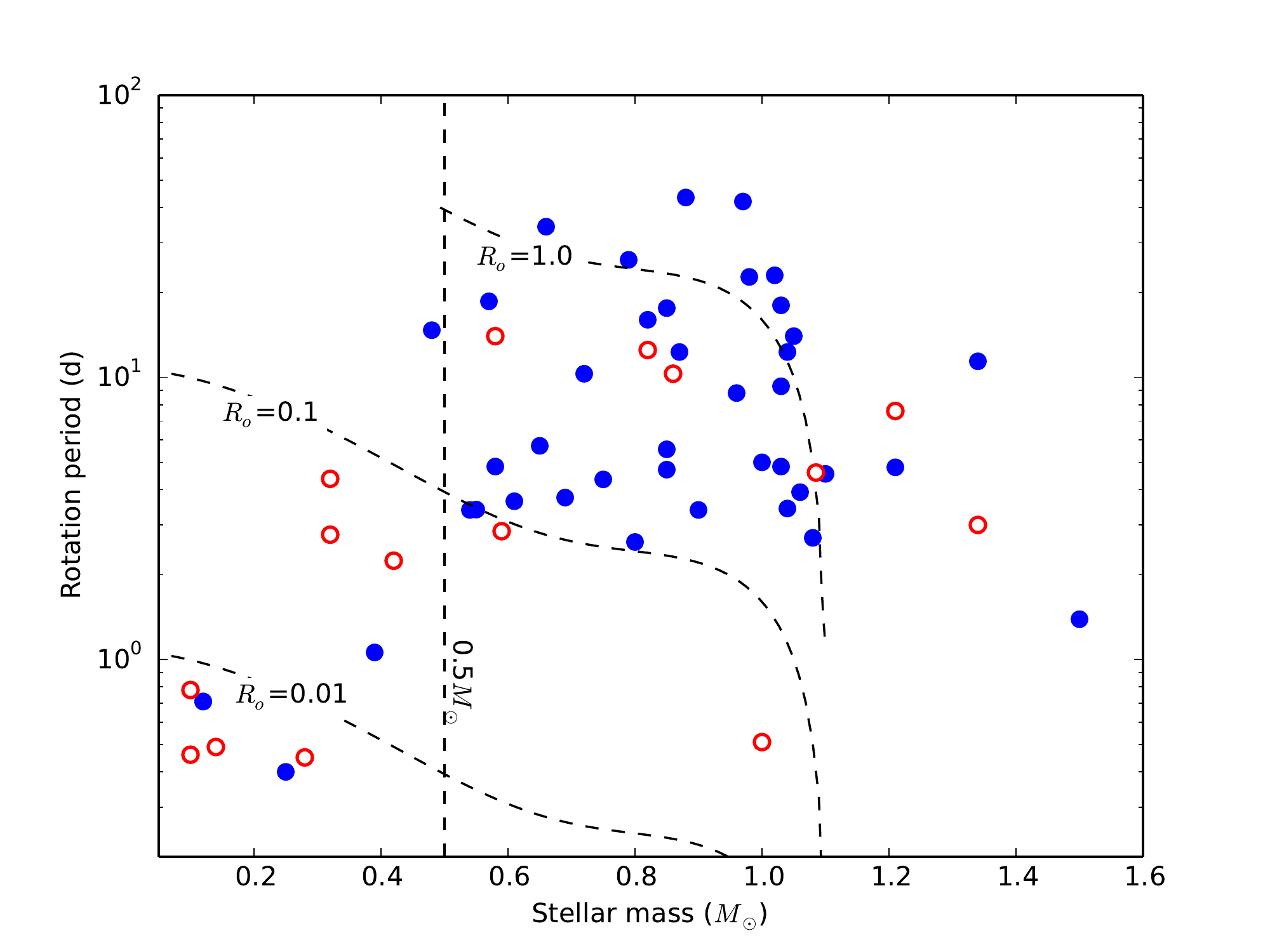}
	\end{center}
	\caption{The rotation periods and masses of each star in our sample. Filled blue points indicate stars observed at one epoch while open red data points indicate stars observed over multiple epochs. Dashed lines are included showing a stellar mass of $0.5\,{\rm M}_\odot$ and Rossby numbers of 0.01, 0.1 and 1.0.}
	\label{fig:paramSpace}
\end{figure}

\section{Stellar sample}
\label{sec:Sample}
For this work, we employ 90 magnetic maps of 55 dwarf stars with spectral types spanning F, G, K and M. The original publications for each map are referenced in Table \ref{tab:StarProperties}. Additionally, parameters for each star are listed in Table \ref{tab:StarProperties}. Fig. \ref{fig:paramSpace} shows the rotation periods and masses of the stars in our sample. Stars with a single map are plotted with filled blue points while stars with multiple maps over many epochs are plotted with open red points. It is worth noting that, though we principally focus on mass and rotation rates in this work, there may be other parameters, such as stellar age and the presence of planets, that affect the magnetic properties of stars. Each map contains information about the large-scale magnetic field at the surface of the star. As outlined by \citet{Donati2006}, the radial, meridional and azimuthal components of the magnetic field of each star ($B_r$,$B_{\theta}$,$B_{\phi}$ respectively) are given by

\begin{equation}
	B_r(\theta,\phi)=\sum_{\ell,m} \alpha_{\ell,m} Y_{\ell,m}(\theta,\phi),
	\label{eq:Br}
\end{equation}

\begin{equation}
	B_{\theta}(\theta,\phi)=-\sum_{\ell,m}[\beta_{\ell,m} Z_{\ell,m}(\theta,\phi) + \gamma_{\ell,m} X_{\ell,m}(\theta,\phi)],
\end{equation}

\begin{equation}
	B_{\phi}(\theta,\phi)=-\sum_{\ell,m}[\beta_{\ell,m} X_{\ell,m}(\theta,\phi) - \gamma_{\ell,m} Z_{\ell,m}(\theta,\phi)],
	\label{eq:Bphi}
\end{equation}
where

\begin{equation}
	Y_{\ell,m}(\theta,\phi)=c_{\ell,m}P_{\ell,m}(\cos\theta)\rm{e}^{\rm{i}m\phi},
\end{equation}

\begin{equation}
	Z_{\ell,m}(\theta,\phi)=\frac{c_{\ell,m}}{\ell+1}\frac{\partial P_{\ell,m}(\cos\theta)}{\partial \theta} \rm{e}^{\rm{i}m\phi},
\end{equation}

\begin{equation}
	X_{\ell,m}(\theta,\phi)=\frac{c_{\ell,m}}{\ell+1} \frac{P_{\ell,m}(\cos\theta)}{\sin \theta} im \rm{e}^{\rm{i}m\phi},
\end{equation}

\begin{equation}
	c_{\ell,m}=\sqrt{\frac{2\ell+1}{4\pi} \frac{(\ell-m)!}{(\ell+m)!}}.
	\label{eq:clm}
\end{equation}
In equations (\ref{eq:Br}) - (\ref{eq:clm}), $P_{\ell,m}(\cos\theta)$ are the associated Legendre polynomials and $\ell$ and $m$ are the order and degree respectively. The large-scale magnetic field of a star can therefore be fully expressed by the $\alpha_{\ell,m}$, $\beta_{\ell,m}$ and $\gamma_{\ell,m}$ coefficients. These are determined by inverting a series of circularly polarised  Stokes V profiles [for a more detailed discussion on this process, see \citet{Donati2006}]. The $\alpha_{\ell,m}$ coefficients characterise the radial component of the field, the $\beta_{\ell,m}$ coefficients characterise the meridional and azimuthal components of the poloidal field and the $\gamma_{\ell,m}$ coefficients characterise the meridional and azimuthal components of the toroidal field. Throughout the rest of this paper, we will make use of the surface averaged quantity, $\langle B^2 \rangle$, where the components of $B$ are given by equation (\ref{eq:Br}) - (\ref{eq:Bphi}). This is proportional to the average magnetic energy density over the surface of a given star.

\begin{figure}
	\begin{center}
	\includegraphics[trim=1cm 1cm 1cm 0.5cm, clip=true, width=\columnwidth]{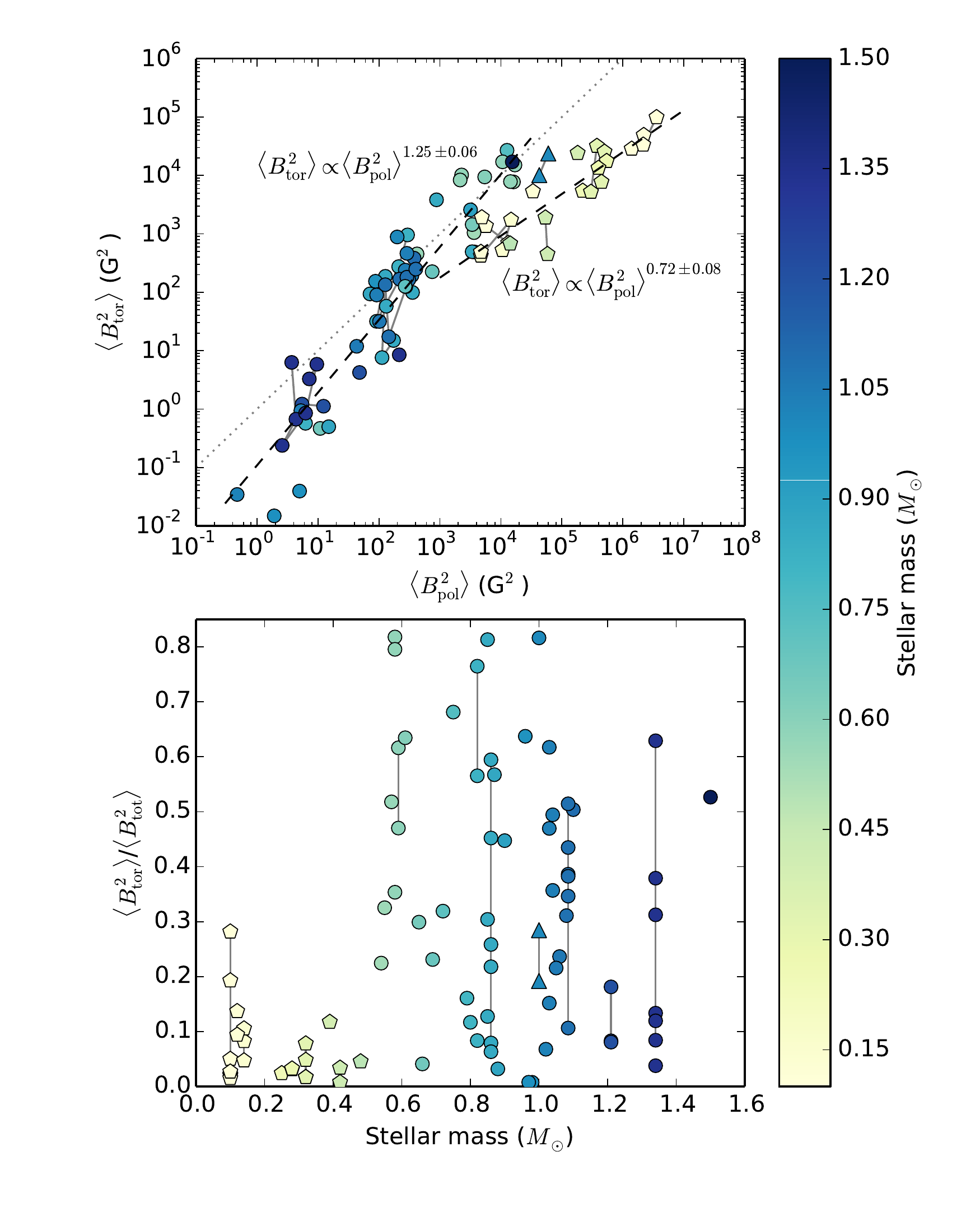}
	\end{center}
	\caption{Top: toroidal magnetic energy against poloidal magnetic energy. Stars with multiple maps are connected by grey lines. The dotted line indicates $\langle B^2_{\rm{tor}}\rangle=\langle B^2_{\rm{pol}}\rangle$. The sample is split into stars less massive (pentagon markers) and more massive than $0.5\,{\rm M}_\odot$ (circle markers). See text for further discussion of how these sub-samples were chosen. The two dashed lines are best fit lines for these sub-samples; $\langle B^2_{\rm{tor}}\rangle \propto \langle B^2_{\rm{pol}}\rangle^{a}$ with $a=0.72\pm 0.08$ and $a=1.25\pm 0.06$ for $\,{\rm M} < 0.5\,{\rm M}_\odot$ and $\,{\rm M} > 0.5\,{\rm M}_\odot$ respectively. AB Dor is shown with triangles. Each point is colour coded by stellar mass. Bottom: toroidal energy fraction against stellar mass. Format is the same as the top panel. Data-points are also colour coded by stellar mass to aid comparison with the top panel. While the two panels show very similar information, the difference in behaviours of the two mass ranges is much clearer in the bottom panel.}
	\label{fig:budget}
\end{figure}

\begin{figure}
	\begin{center}
	\includegraphics[trim=1cm 1cm 1cm 0.5cm, clip=true, width=\columnwidth]{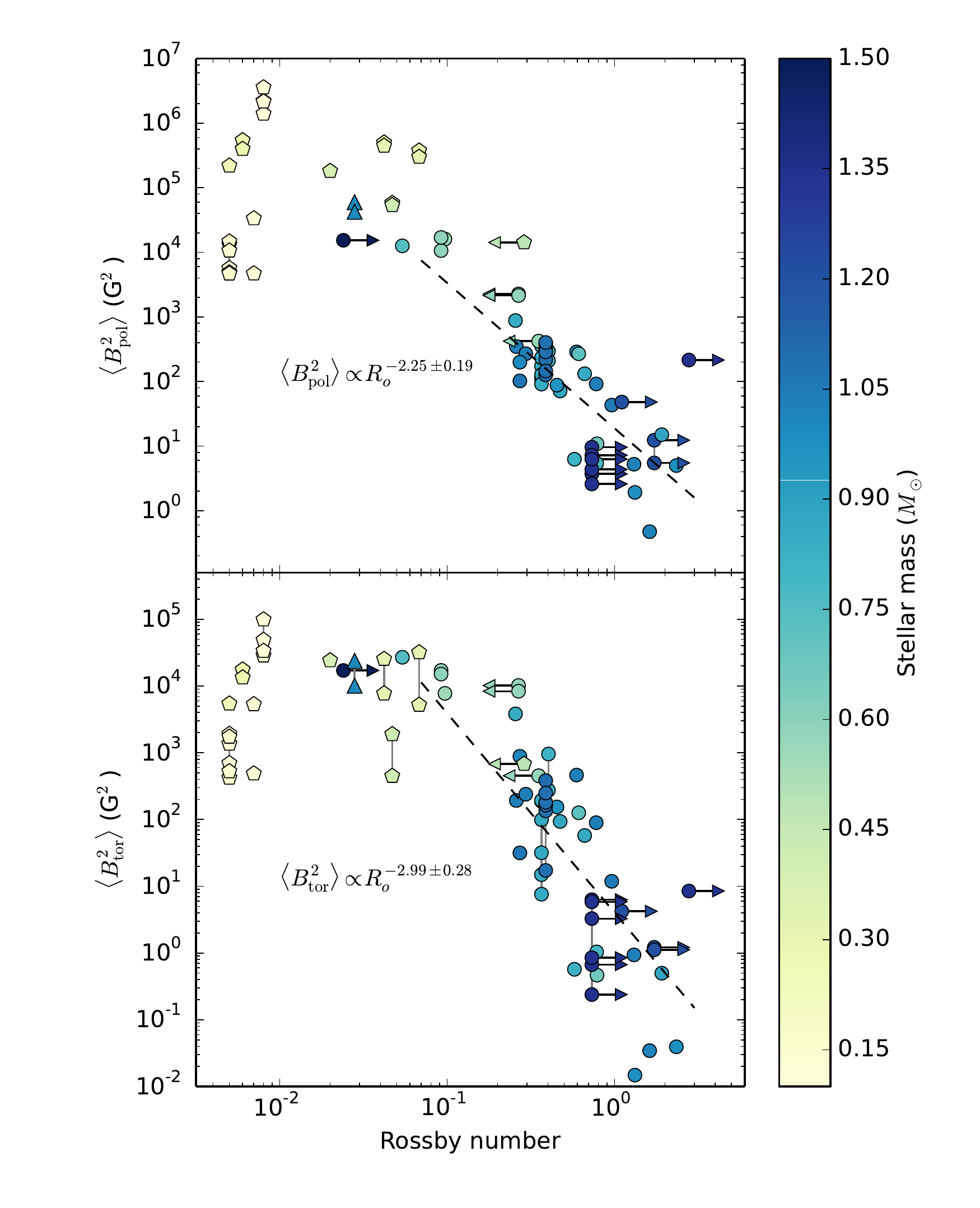}
	\end{center}
	\caption{Poloidal (top) and toroidal (bottom) magnetic energy against Rossby number. The formatting is the same as Fig. \ref{fig:budget}. Right/left facing arrows indicate stars that only have lower/upper estimates for their Rossby numbers. The saturated and unsaturated regimes can be clearly seen with the transition occurring at a Rossby number of approximately 0.1. Fits to the stars in the unsaturated regime, $\langle B^2_{\mathrm{pol}}\rangle \propto R_o^{-2.25 \pm 0.19}$ and $\langle B^2_{\mathrm{tor}}\rangle \propto R_o^{-2.99 \pm 0.28}$, are shown with dashed lines. Note: the magnetic energy axes of the two plots are not the same.}
	\label{fig:sat}
\end{figure}

\section{Results}
\label{sec:Results}
\subsection{Magnetic energy budget and saturation}
\label{subsec:budget}
One of the principle ways to split the stellar magnetic energy is into its poloidal and toroidal components which is shown in Fig. \ref{fig:budget}. The top panel shows toroidal magnetic energy density against poloidal magnetic energy density while the bottom panel shows the toroidal energy fraction against stellar mass. Both panels are colour coded by stellar mass aiding comparison between them. 

The top panel shows that the toroidal energy is an increasing function of poloidal energy. We can attempt to fit a power law of the form $\langle B^2_{\rm{tor}}\rangle \propto \langle B^2_{\rm{pol}}\rangle^{a}$. However, the sample seems to consist of two sub-samples. The stars with higher magnetic energy densities appear to have a smaller power index, $a$, than the lower energy stars. A priori, it is not clear which stars should be included in which sub-sample. In the bottom panel, a change of behaviour is evident at approximately 0.5$\,{\rm M}_\odot$. Stars with a larger mass than this can have large toroidal energy fractions but lower mass stars cannot. Though the two panels show essentially the same information, this break in behaviour at 0.5$\,{\rm M}_\odot$ is much clearer in the bottom panel. A number of authors have previously discussed a sudden change in the magnetic properties of M dwarfs at roughly $0.5\,{\rm M}_\odot$ \citep{Donati2008a,Morin2008,Morin2010,Gregory2012}. They note that this break is roughly coincident with the fully convective limit suggesting a link with the change in internal structure. Dividing our sample on this basis, we find power index values of $a=0.72\pm 0.08$ and $a=1.25\pm 0.06$ for stars less and more massive than 0.5$\,{\rm M}_\odot$ respectively. These power laws are plotted in the top panel with dashed lines. It is worth noting that, among the $\rm{M} <0.5\,{\rm M}_\odot$ stars, it is the dipole dominate stars that deviate the most from the higher index power law. The non-dipolar stars in the bistable regime, as discussed by \citet{Morin2010}, are roughly compatible with the other power law. Additionally, theoretical models predict that these non-dipolar stars can vary cyclically and are able to generate significant toroidal fields \citep[e.g.][]{Gastine2013}.

\begin{table*}
\begin{minipage}{190mm}
	\caption{Parameters of our sample: star ID, alternative name, stellar mass, rotation period, Rossby number, $\langle B^2 \rangle$ (which is proportional to the magnetic energy density), toroidal magnetic energy (as a percentage of total energy), axisymmetric magnetic energy (as a percentage of total energy), poloidal axisymmetric magnetic energy (as a percentage of poloidal energy), toroidal axisymmetric magnetic energy (as a percentage of toroidal energy) and the observation epoch. Similarly to \citet{Vidotto2014}, we have grouped the stars into solar-like stars, young Suns, hot Jupiter hosts and M dwarfs. References indicate the paper where the magnetic map was originally published. For the remaining parameters, references can be found in \citet{Vidotto2014} and references therein.}
	\label{tab:StarProperties}
	\begin{tabular}{lccccccccccc}
		\hline
		Star & Alt. & $M_{\star}$ & $P_{\rm{rot}}$ & $R_o$ & $\log \left[\frac{\langle B^2 \rangle}{\rm{G}^2}\right]$ & Tor & Axi & Pol Axi & Tor Axi & Obs & Ref.\\
		ID & name & [$\,{\rm M}_\odot$] & [d] & & & [\% total] & [\%total] & [\%pol] & [\%tor] & epoch & \\
		\hline
\textbf{Solar-like} \\																							
\textbf{stars} \\																							
HD 3651	&		&	0.88	&	43.4	&	1.916	&	1.19	&	3	&	87	&	87	&	98	&	-	&	\citet{Petit2015}	\\
HD 9986	&		&	1.02	&	23	&	1.621	&	-0.29	&	7	&	53	&	50	&	94	&	-	&	\citet{Petit2015}	\\
HD 10476	&		&	0.82	&	16	&	0.576	&	0.83	&	8	&	4	&	0	&	40	&	-	&	\citet{Petit2015}	\\
HD 20630	&	$\kappa$ Ceti 	&	1.03	&	9.3	&	0.593	&	2.88	&	62	&	66	&	30	&	89	&	2012 Oct	&	\citet{Nascimento2014}	\\
HD 22049	&	$\epsilon$ Eri	&	0.86	&	10.3	&	0.366	&	2.27	&	8	&	16	&	10	&	92	&	2007 Jan	&	\citet{Jeffers2014}	\\
...	&	...	&	...	&	...	&	...	&	2.08	&	6	&	58	&	59	&	45	&	2008 Jan	&	\citet{Jeffers2014}	\\
...	&	...	&	...	&	...	&	...	&	2.50	&	59	&	72	&	36	&	96	&	2010 Jan	&	\citet{Jeffers2014}	\\
...	&	...	&	...	&	...	&	...	&	2.09	&	26	&	63	&	55	&	86	&	2011 Oct	&	\citet{Jeffers2014}	\\
...	&	...	&	...	&	...	&	...	&	2.63	&	45	&	59	&	43	&	80	&	2012 Oct	&	\citet{Jeffers2014}	\\
...	&	...	&	...	&	...	&	...	&	2.66	&	22	&	36	&	21	&	91	&	2013 Sep	&	\citet{Jeffers2014}	\\
HD 39587	&		&	1.03	&	4.83	&	0.295	&	2.71	&	47	&	36	&	4	&	72	&	-	&	\citet{Petit2015}	\\
HD 56124	&		&	1.03	&	18	&	1.307	&	0.79	&	15	&	90	&	90	&	91	&	-	&	\citet{Petit2015}	\\
HD 72905	&		&	1	&	5	&	0.272	&	3.03	&	82	&	81	&	13	&	97	&	-	&	\citet{Petit2015}	\\
HD 73350	&		&	1.04	&	12.3	&	0.777	&	2.26	&	49	&	44	&	0	&	90	&	-	&	\citet{Petit2015}	\\
HD 75332	&		&	1.21	&	4.8	&	$>$1.105	&	1.72	&	8	&	73	&	76	&	37	&	-	&	\citet{Petit2015}	\\
HD 78366	&		&	1.34	&	11.4	&	$>$2.781	&	2.35	&	4	&	91	&	93	&	29	&	-	&	\citet{Petit2015}	\\
HD 101501	&		&	0.85	&	17.6	&	0.663	&	2.28	&	30	&	42	&	25	&	80	&	-	&	\citet{Petit2015}	\\
HD 131156A	&	$\xi$ Boo A	&	0.85	&	5.56	&	0.256	&	3.67	&	81	&	81	&	10	&	97	&	-	&	\citet{Petit2015}	\\
HD 131156B	&	$\xi$ Boo B	&	0.72	&	10.3	&	0.611	&	2.60	&	32	&	42	&	24	&	81	&	-	&	\citet{Petit2015}	\\
HD 146233	&	18 Sco	&	0.98	&	22.7	&	1.324	&	0.29	&	1	&	9	&	9	&	5	&	2007 Aug	&	\citet{Petit2008}	\\
HD 166435	&		&	1.04	&	3.43	&	0.259	&	2.73	&	36	&	49	&	37	&	70	&	-	&	\citet{Petit2015}	\\
HD 175726	&		&	1.06	&	3.92	&	0.272	&	2.13	&	24	&	25	&	12	&	68	&	-	&	\citet{Petit2015}	\\
HD 190771	&		&	0.96	&	8.8	&	0.453	&	2.38	&	64	&	75	&	34	&	98	&	-	&	\citet{Petit2015}	\\
HD 201091A	&	61 Cyg A	&	0.66	&	34.2	&	0.786	&	1.05	&	4	&	59	&	58	&	96	&	-	&	\citet{Petit2015}	\\
HD 206860	&	HN Peg	&	1.1	&	4.55	&	0.388	&	2.88	&	50	&	66	&	40	&	92	&	-	&	\citet{Petit2015}	\\
...	&	...	&	...	&	...	&	...	&	2.59	&	43	&	69	&	52	&	92	&	2007 Jul	&	\citet{Saikia2015}	\\
...	&	...	&	...	&	...	&	...	&	2.42	&	51	&	54	&	24	&	82	&	2008 Aug	&	\citet{Saikia2015}	\\
...	&	...	&	...	&	...	&	...	&	2.21	&	11	&	45	&	48	&	28	&	2009 Jun	&	\citet{Saikia2015}	\\
...	&	...	&	...	&	...	&	...	&	2.67	&	35	&	44	&	38	&	53	&	2010 Jul	&	\citet{Saikia2015}	\\
...	&	...	&	...	&	...	&	...	&	2.67	&	39	&	38	&	17	&	72	&	2011 Jul	&	\citet{Saikia2015}	\\
...	&	...	&	...	&	...	&	...	&	2.81	&	38	&	77	&	69	&	90	&	2013 Jul	&	\citet{Saikia2015}	\\
\textbf{Young} \\																							
\textbf{Suns} \\																							
Ab Doradus	&		&	0.76	&	0.5	&	0.026	&	4.92	&	28	&	18	&	9	&	41	&	2001 Dec	&	\citet{Donati2003Dynamo}	\\
...	&		&	...	&	...	&	...	&	4.72	&	19	&	12	&	7	&	35	&	2002 Dec	&	\citet{Donati2003Dynamo}	\\
BD-16 351	&		&	0.9	&	3.39	&	-	&	3.76	&	45	&	41	&	3	&	88	&	2012 Sep	&	\citet{FolsomToupies}	\\
HD 106506	&		&	1.5	&	1.39	&	$>$0.024	&	4.51	&	53	&	47	&	17	&	75	&	2007 Apr	&	\citet{Waite2011}	\\
HII 296	&		&	0.8	&	2.61	&	-	&	3.62	&	12	&	36	&	35	&	43	&	2009 Oct	&	\citet{FolsomToupies}	\\
HII 739	&		&	1.08	&	2.7	&	-	&	2.60	&	31	&	33	&	20	&	60	&	2009 Oct	&	\citet{FolsomToupies}	\\
HIP 12545	&		&	0.58	&	4.83	&	-	&	4.35	&	35	&	60	&	47	&	83	&	2012 Sep	&	\citet{FolsomToupies}	\\
HIP 76768	&		&	0.61	&	3.64	&	-	&	4.17	&	63	&	86	&	73	&	94	&	2013 May	&	\citet{FolsomToupies}	\\
TYC0486-4943-1	&		&	0.69	&	3.75	&	-	&	2.99	&	23	&	25	&	12	&	67	&	2013 Jun	&	\citet{FolsomToupies}	\\
TYC5164-567-1	&		&	0.85	&	4.71	&	-	&	3.59	&	13	&	54	&	56	&	35	&	2013 Jun	&	\citet{FolsomToupies}	\\
TYC6349-0200-1	&		&	0.54	&	3.39	&	-	&	3.67	&	22	&	32	&	24	&	61	&	2013 Jun	&	\citet{FolsomToupies}	\\
TYC6878-0195-1	&		&	0.65	&	5.72	&	-	&	3.68	&	30	&	38	&	19	&	83	&	2013 Jun	&	\citet{FolsomToupies}	\\

	\hline
\end{tabular}
\end{minipage}
\end{table*}

\begin{table*}
\begin{minipage}{190mm}
	\contcaption{}
	\begin{tabular}{lcccccccccccc}
		\hline
		Star & Alt. & $M_{\star}$ & $P_{\rm{rot}}$ & $R_o$ & $\log \left[\frac{\langle B^2 \rangle}{\rm{G}^2}\right]$ & Tor & Axi & Pol Axi & Tor Axi & Obs & Ref.\\
		ID & name & [$\,{\rm M}_\odot$] & [d] & & & [\% total] & [\%total] & [\%pol] & [\%tor] & epoch & \\
		\hline
\textbf{Hot Jupiter} \\									
\textbf{hosts} \\			
$\tau$ Boo	&		&	1.34	&	3	&	$>$0.732	&	1.00	&	63	&	63	&	17	&	90	&	2008 Jan	&	\citet{Fares2009}	\\
...	&		&	...	&	...	&	...	&	0.70	&	8	&	57	&	59	&	37	&	2008 Jun	&	\citet{Fares2009}	\\
...	&		&	...	&	...	&	...	&	0.45	&	13	&	25	&	23	&	37	&	2008 Jul	&	\citet{Fares2009}	\\
...	&		&	...	&	...	&	...	&	0.85	&	12	&	33	&	35	&	15	&	2009 May	&	\citet{Fares2013}	\\
...	&		&	...	&	...	&	...	&	1.19	&	38	&	25	&	18	&	37	&	2010 Jan	&	\citet{Fares2013}	\\
...	&		&	...	&	...	&	...	&	1.02	&	31	&	38	&	28	&	59	&	2011 Jan	&	\citet{Fares2013}	\\
HD 46375	&		&	0.97	&	42	&	2.34	&	0.70	&	1	&	77	&	77	&	86	&	2008 Jan	&	\citet{Fares2013}	\\
HD 73256	&		&	1.05	&	14	&	0.962	&	1.74	&	22	&	20	&	3	&	79	&	2008 Jan	&	\citet{Fares2013}	\\
HD 102195	&		&	0.87	&	12.3	&	0.473	&	2.22	&	57	&	60	&	23	&	88	&	2008 Jan	&	\citet{Fares2013}	\\
HD 130322	&		&	0.79	&	26.1	&	0.782	&	0.81	&	16	&	64	&	58	&	96	&	2008 Jan	&	\citet{Fares2013}	\\
HD 179949	&		&	1.21	&	7.6	&	$>$1.726	&	0.83	&	18	&	58	&	54	&	78	&	2007 Jun	&	\citet{Fares2012}	\\
...	&		&	...	&	...	&	...	&	1.13	&	8	&	34	&	34	&	32	&	2009 Sep	&	\citet{Fares2012}	\\
HD 189733	&		&	0.82	&	12.5	&	0.403	&	2.69	&	57	&	62	&	26	&	91	&	2007 Jun	&	\citet{Fares2010}	\\
...	&		&	...	&	...	&	...	&	3.10	&	76	&	77	&	16	&	96	&	2008 Jul	&	\citet{Fares2010}	\\
\textbf{M dwarf}\\
\textbf{stars}\\		
GJ 569A	&	CE Boo	&	0.48	&	14.7	&	$<$0.288	&	4.17	&	5	&	96	&	96	&	100	&	2008 Jan	&	\citet{Donati2008a}	\\
GJ 410	&	DS Leo	&	0.58	&	14	&	$<$0.267	&	4.10	&	82	&	92	&	58	&	99	&	2007 Jan	&	\citet{Donati2008a}	\\
...	&	...	&	...	&	...	&	...	&	4.02	&	80	&	78	&	15	&	94	&	2007 Dec	&	\citet{Donati2008a}	\\
GJ 182	&		&	0.75	&	4.35	&	0.054	&	4.60	&	68	&	66	&	15	&	90	&	2007 Jan	&	\citet{Donati2008a}	\\
GJ 49	&		&	0.57	&	18.6	&	$<$0.352	&	2.94	&	52	&	84	&	67	&	100	&	2007 Jul	&	\citet{Donati2008a}	\\
GJ 388	&	AD Leo	&	0.42	&	2.24	&	0.047	&	4.77	&	1	&	96	&	97	&	22	&	2007 Jan	&	\citet{Morin2008}	\\
...	&	...	&	...	&	...	&	...	&	4.75	&	3	&	89	&	92	&	8	&	2008 Jan	&	\citet{Morin2008}	\\
GJ 494A	&	DT Vir	&	0.59	&	2.85	&	0.092	&	4.44	&	62	&	61	&	12	&	91	&	2007 Jan	&	\citet{Donati2008a}	\\
...	&	...	&	...	&	...	&	...	&	4.51	&	47	&	49	&	17	&	85	&	2007 Dec	&	\citet{Donati2008a}	\\
GJ 896 A	&	EQ Peg A	&	0.39	&	1.06	&	0.02	&	5.31	&	12	&	66	&	71	&	29	&	2006 Aug	&	\citet{Morin2008}	\\
GJ 896 B	&	EQ Peg B	&	0.25	&	0.4	&	0.005	&	5.35	&	2	&	92	&	93	&	42	&	2006 Aug	&	\citet{Morin2008}	\\
GJ 873	&	EV Lac	&	0.32	&	4.37	&	0.068	&	5.61	&	8	&	31	&	28	&	60	&	2006 Aug	&	\citet{Morin2008}	\\
...	&	...	&	...	&	...	&	...	&	5.48	&	2	&	29	&	29	&	18	&	2007 July	&	\citet{Morin2008}	\\
GJ 1111	&	DX Cnc	&	0.1	&	0.46	&	0.005	&	4.14	&	5	&	77	&	77	&	65	&	2007	&	\citet{Morin2010}	\\
...	&	...	&	...	&	...	&	...	&	3.85	&	19	&	40	&	30	&	82	&	2008	&	\citet{Morin2010}	\\
...	&	...	&	...	&	...	&	...	&	3.83	&	28	&	65	&	61	&	77	&	2009	&	\citet{Morin2010}	\\
GJ 1156	&		&	0.14	&	0.49	&	0.005	&	3.71	&	8	&	3	&	2	&	18	&	2007	&	\citet{Morin2010}	\\
...	&		&	...	&	...	&	...	&	4.21	&	11	&	17	&	12	&	57	&	2008	&	\citet{Morin2010}	\\
...	&		&	...	&	...	&	...	&	4.05	&	5	&	1	&	1	&	7	&	2009	&	\citet{Morin2010}	\\
GJ 1245B	&		&	0.12	&	0.71	&	0.007	&	4.59	&	14	&	10	&	6	&	37	&	2006	&	\citet{Morin2010}	\\
...	&		&	...	&	...	&	...	&	3.72	&	9	&	15	&	13	&	30	&	2008	&	\citet{Morin2010}	\\
GJ 9520	&	OT Ser	&	0.55	&	3.4	&	0.097	&	4.38	&	33	&	70	&	61	&	90	&	2008 Feb	&	\citet{Donati2008a}	\\
V374 Peg	&		&	0.28	&	0.45	&	0.006	&	5.75	&	3	&	79	&	81	&	5	&	2005 Aug	&	\citet{Morin2008b}	\\
...	&		&	...	&	...	&	...	&	5.61	&	3	&	76	&	79	&	2	&	2006 Aug	&	\citet{Morin2008b}	\\
GJ 412B	&	WX UMa	&	0.1	&	0.78	&	0.008	&	6.16	&	2	&	92	&	92	&	89	&	2006	&	\citet{Morin2010}	\\
...	&	...	&	...	&	...	&	...	&	6.35	&	2	&	92	&	94	&	38	&	2007	&	\citet{Morin2010}	\\
...	&	...	&	...	&	...	&	...	&	6.34	&	2	&	84	&	86	&	3	&	2008	&	\citet{Morin2010}	\\
...	&	...	&	...	&	...	&	...	&	6.56	&	3	&	94	&	96	&	20	&	2009	&	\citet{Morin2010}	\\
GJ 285	&	YZ Cmi	&	0.32	&	2.77	&	0.04	&	5.72	&	5	&	57	&	59	&	19	&	2007 Jan	&	\citet{Morin2008}	\\
...	&	...	&	...	&	...	&	...	&	5.66	&	2	&	87	&	88	&	25	&	2008 Jan	&	\citet{Morin2008}	\\
		\hline
	\end{tabular}
\end{minipage}
\end{table*}

As a brief aside, we discuss the possibility that the smaller toroidal energy fraction of stars less massive than 0.5$\,{\rm M}_\odot$ is a result of the ZDI technique. ZDI captures the large-scale fields but is insensitive to small-scale fields due to flux cancellation effects. \citet{Reiners2009} show that the majority of magnetic flux may be missed when the stellar field is reconstructed using only the Stokes V signal when compared to the Stokes I signal. If the lowest mass M dwarfs have a large fraction of their magnetic energy stored in small scale fields, in the form of star spots for example, these fields may not be reconstructed by the ZDI technique. However, there is no reason to expect ZDI to preferentially miss toroidal field over poloidal field.

It is a well known observational result that the ratio of X-ray to bolometric luminosity, $R_{\textrm{X}}=L_{\textrm{X}}/L_{\textrm{bol}}$, increases with decreasing Rossby number, $R_o$, and saturates for stars with Rossby numbers less than approximately 0.1 \citep{Pizzolato2003,Wright2011}. Here, the Rossby number is defined as the ratio of the stellar rotation period to the convective turnover time. \citet{Vidotto2014} showed that stellar magnetism has the same qualitative dependence on Rossby number. However, this result was derived using the radial component of the surface fields only, and hence does not consider the toroidal field. In Fig. \ref{fig:sat}, we plot the poloidal (top panel) and toroidal (bottom panel) magnetic energy densities as a function of Rossby number. We use the same Rossby number estimates as \citet{Vidotto2014}, where further discussion of the estimates can be found. Left and right facing arrows indicate stars where the Rossby number estimate is only an upper or lower limit respectively.

It is clear to see that both components qualitatively follow the well known behaviour with some quantitative differences. In both cases, the cutoff between the saturated and unsaturated regimes occurs at $R_o \sim 0.1$. However, in the saturated regime, the average magnetic energy of the poloidal fields is higher than that of the toroidal fields by just over an order of magnitude. Additionally, in the unsaturated regime, the slope is steeper for the toroidal component. The similar behaviour indicates that the same mechanism is responsible for generating both components or that both components are generated from each other.

\citet{Wright2011} suggested that the difference in $R_X$ behaviour in the saturated and unsaturated regimes can be attributed to different dynamo mechanisms operating in each of the regimes rather than any actual saturation effect. They argued this on the basis that the age at which stars transition from the rotational C sequence to the I sequence \citep{Barnes2003} is coincident with the transition from saturated to unsaturated regimes at $R_o \sim 0.1$ (see their Fig. 4). It is interesting to note that the majority of the stars in the saturated regime in Fig. \ref{fig:sat} have $\,{\rm M} < 0.5\,{\rm M}_\odot$ (pentagon symbols). Assuming that the unsaturated stars are those more massive than $0.5\,{\rm M}_\odot$, we find fits of $\langle B^2_{\rm{pol}}\rangle \propto R_o^{-2.25 \pm 0.19}$ and $\langle B^2_{\rm{tor}}\rangle \propto R_o^{-2.99 \pm 0.28}$ for the poloidal and toroidal fields respectively. These fits are plotted with dashed lines. It is worth noting that these two fits and the fit in Fig. \ref{fig:budget} with the higher power index, $a=1.25$, are the three possible 2D projections of the same relatively tight sequence of stars in $(\langle B^2_{\rm{pol}}\rangle, \langle B^2_{\rm{tor}}\rangle, R_o)$ parameter space.

If there is a physical basis for the saturated stars having masses less than $0.5\,{\rm M}_\odot$, the fact that they appear to have a different power index, $a$, to the rest of the stars, in Fig. \ref{fig:budget}, is further evidence for the suggestion of \citet{Wright2011}. However, care must be taken with this interpretation. Fig. \ref{fig:paramSpace} shows that the majority of stars in our sample that are less than $0.5\,{\rm M}_\odot$ also have Rossby numbers less than 0.1. Therefore, our interpretation is not that stars in the saturated regime must be less massive than $0.5\,{\rm M}_\odot$. Rather, it is that stars in the saturated regime are those with Rossby numbers less than $\sim$0.1 which also happen to be stars less massive then $0.5\,{\rm M}_\odot$ in our sample.  Noticeably, with the exception of AB Dor ($R_o = 0.028$, $\rm{M} = 1.0{\rm M}_\odot$), there is a dearth of stars with Rossby numbers less than $\sim$0.1 and masses bigger than $0.5\,{\rm M}_\odot$ in our sample. That the lowest mass stars in our sample are also the fastest rotators is not surprising since these stars spin down less rapidly than higher mass stars \citep{Mohanty2003}. Correspondingly $R_o  \lesssim  0.1$, $\rm{M}_{\star}>0.5\,{\rm M}_\odot$ stars are harder to find and map. Indeed a number of authors have previously commented on the difficulty in separating stellar mass and Rossby number effects due to this bias \citep{Donati2008a,Morin2008,Morin2010,Reiners2009,Gastine2013}. Even when we consider AB Dor (plotted with triangles in Figs. \ref{fig:budget} and \ref{fig:sat}), which does fall into the $R_o \lesssim  0.1$, $\rm{M}_{\star}>0.5\,{\rm M}_\odot$ region of parameter space, the picture does not become any clearer. In the top panel of Fig. \ref{fig:budget}, AB Dor lies relatively close to both sub-samples such that it is difficult to tell which sub-sample it would be more appropriate for it to be in. Likewise, in Fig. \ref{fig:sat}, AB Dor falls in an intermediate region. On one hand, one might consider it a saturated star on the basis of its Rossby number which is smaller than $R_o \sim 0.1$, the value typically used to delineate the saturated and unsaturated regimes. However, one might also consider it to be an unsaturated star on the basis that it follows the same trend as the other unsaturated stars, lying at the tail end of that sequence. If the hypothesis of \citet{Wright2011} is correct, one might expect the magnetic fields of the stars in this region of parameter space to obey the power law with the smaller index from Fig. \ref{fig:budget}. 

A potential problem with this interpretation is that our two sub-samples were initially chosen on the basis of stellar mass, since this parameter better discriminates between the two power laws in Fig. \ref{fig:budget}, rather than Rossby number. It may be the case that the two power laws in Fig. \ref{fig:budget} do not correspond to the saturated and unsaturated regimes. Indeed, until more stars in the $R_o<0.1$, $M_{\star}>0.5\,{\rm M}_\odot$ region of parameter space (corresponding to stars of rotation periods of less than a few days - see Fig. \ref{fig:paramSpace}) have their surface fields mapped, it will be difficult to conclusively confirm or reject this hypothesis. 

Recently, \citet{Reiners2014} reinterpreted the data of \citet{Wright2011}. These authors show that, in the unsaturated regime, $R_{\textrm{X}}=L_{\textrm{X}}/L_{\textrm{bol}}$ shows less scatter when plotted against $R_\ast^{-4}P_{\rm{rot}}^{-2}$, where $R_\ast$ and $P_{\rm rot}$ are the stellar radius and rotation period respectively, rather than Rossby number, as is traditional in these types of studies. This formulation is approximately equivalent to $L_{\textrm{X}} \propto P_{\rm{rot}}^{-2}$, i.e. that X-ray luminosity depends only on the stellar rotation period, in the unsaturated regime. The authors make no claims as to whether $R_{\textrm{X}}$ as a function of $R_o$ or $L_{\textrm{X}}$ as a function of $P_{\rm rot}$ is the more physically fundamental relationship. When plotting $\langle B^2_{\rm{pol}}\rangle$ and $\langle B^2_{\rm{tor}}\rangle$ against $R_\ast^{-4}P_{\rm{rot}}^{-2}$ (not shown), we once again find that the data separates into saturated and unsaturated regimes. Similarly to Fig. \ref{fig:sat}, the stars in the saturated regime are those that are less massive than $0.5\,{\rm M}_\odot$. The suggestion that different dynamo mechanisms are present in the saturated and unsaturated regimes is therefore possible under either interpretation of the rotation-activity relation.

\begin{figure}
	\begin{center}
	\includegraphics[trim=0.5cm 0.4cm 1cm 0.5cm, clip=true, width=\columnwidth]{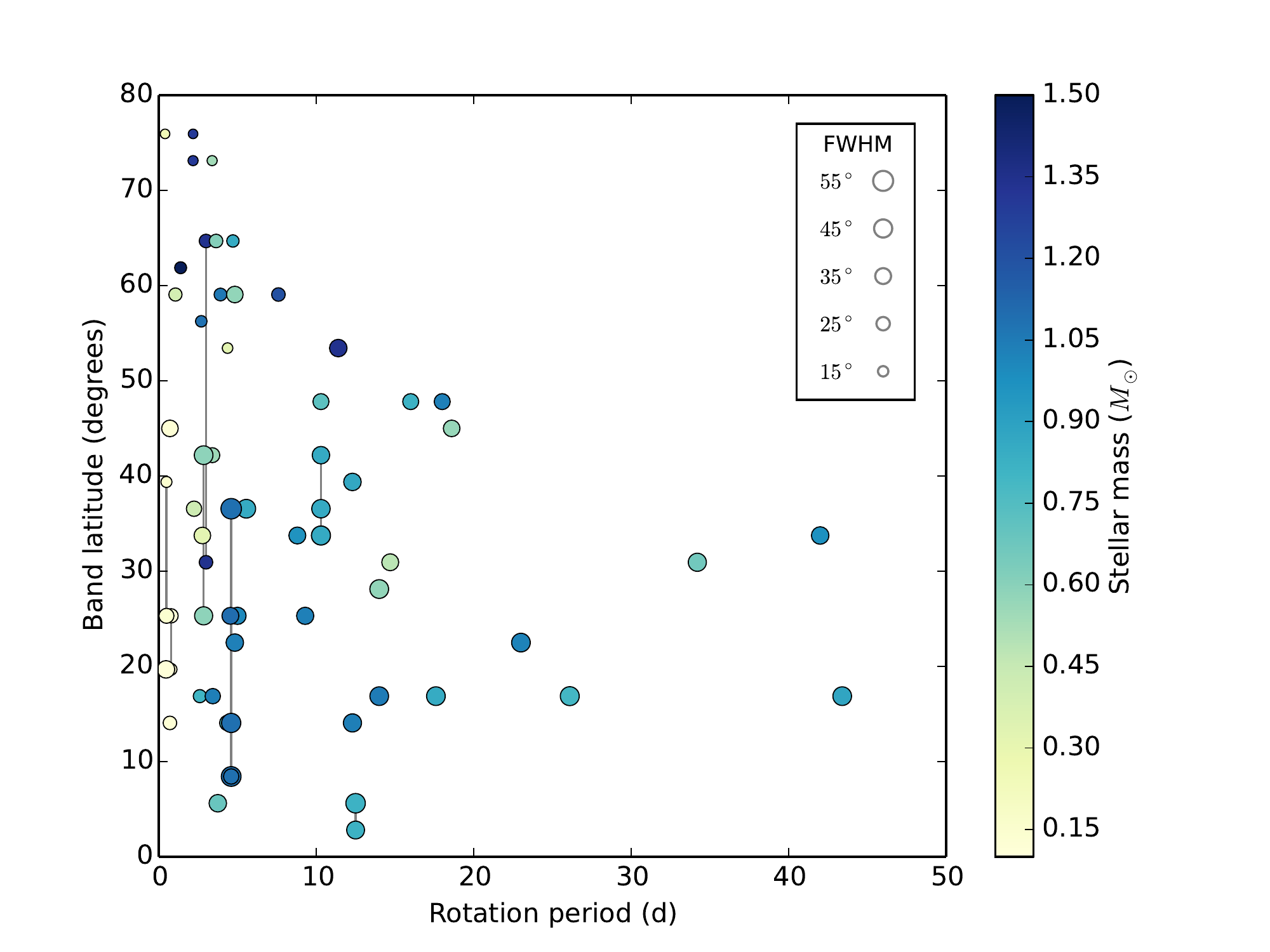}
	\end{center}
	\caption{The latitude at which the toroidal azimuthal field peaks for each star as a function of stellar rotation period. Each point is colour coded by stellar mass. The size of the points indicates the full-width half-maximum value for the band (see inset). Stars with multiple maps are connected by grey lines. Numerical values for the data points are included in Table \ref{tab:latProperties}.}
	\label{fig:bandlat}
\end{figure}

\begin{table*}
\begin{minipage}{160mm}
	\caption{Numerical values for Fig. \ref{fig:bandlat}. For each star the rotation period, latitude of the band and the full-width half-maximum of the band are listed.}
	\label{tab:latProperties}
	\begin{tabular}{lccclccc}
		\hline
		Star & $P_{\rm{rot}}$ & Latitude & FWHM & \hspace{1cm} Star & $P_{\rm{rot}}$ & Latitude & FWHM \\
		ID & [d] & [degrees] & [degrees] & \hspace{1cm} ID & [d] & [degrees] & [degrees]\\
		\hline
HD 3651	&	43.4	&	16.9	&	47.8	&	\hspace{1cm}	TYC5164-567-1	&	4.71	&	64.7	&	21.1	\\
HD 9986	&	23	&	22.5	&	47.8	&	\hspace{1cm}	TYC6349-0200-1	&	3.39	&	73.1	&	14.1	\\
HD 10476	&	16	&	47.8	&	35.2	&	\hspace{1cm}	$\tau$ Boo (2009 May)	&	3	&	64.7	&	25.3	\\
HD 20630	&	9.3	&	25.3	&	40.8	&	\hspace{1cm}	$\tau$ boo (2010 Jan)	&	3	&	30.9	&	25.3	\\
HD 22049 (2007 Jan)	&	10.3	&	42.2	&	42.2	&	\hspace{1cm}	HD 46375	&	42	&	33.8	&	42.2	\\
HD 22049 (2010 Jan)	&	10.3	&	33.8	&	49.2	&	\hspace{1cm}	HD 73256	&	14	&	16.9	&	47.8	\\
HD 22049 (2011 Oct)	&	10.3	&	36.6	&	42.2	&	\hspace{1cm}	HD 102195	&	12.3	&	39.4	&	42.2	\\
HD 22049 (2012 Oct)	&	10.3	&	33.8	&	49.2	&	\hspace{1cm}	HD 130322	&	26.1	&	16.9	&	49.2	\\
HD 22049 (2013 Oct)	&	10.3	&	36.6	&	46.4	&	\hspace{1cm}	HD 179949 (2009 Sep)	&	7.6	&	59.1	&	25.3	\\
HD 39587	&	4.83	&	22.5	&	42.2	&	\hspace{1cm}	HD 189733 (2007 Jun)	&	12.5	&	2.8	&	43.6	\\
HD 56124	&	18	&	47.8	&	35.2	&	\hspace{1cm}	HD 189733 (2008 Jul)	&	12.5	&	5.6	&	52	\\
HD 72905	&	5	&	25.3	&	40.8	&	\hspace{1cm}	GJ 569A	&	14.7	&	30.9	&	39.4	\\
HD 73350	&	12.3	&	14.1	&	45	&	\hspace{1cm}	GJ 410 (2007 Jan)	&	14	&	28.1	&	47.8	\\
HD 78366	&	11.4	&	53.4	&	42.2	&	\hspace{1cm}	GJ 410 (2007 Dec)	&	14	&	28.1	&	46.4	\\
HD 101501	&	17.6	&	16.9	&	47.8	&	\hspace{1cm}	GJ 182	&	4.35	&	14.1	&	32.3	\\
HD 131156A	&	5.56	&	36.6	&	47.8	&	\hspace{1cm}	GJ 49	&	18.6	&	45	&	38	\\
HD 131156B	&	10.3	&	47.8	&	35.2	&	\hspace{1cm}	GJ 388 (2007 Jan)	&	2.24	&	36.6	&	32.3	\\
HD 166435	&	3.43	&	16.9	&	32.3	&	\hspace{1cm}	GJ 494A (2007 Jan)	&	2.85	&	25.3	&	45	\\
HD 175726	&	3.92	&	59.1	&	22.5	&	\hspace{1cm}	GJ 494A (2007 Dec)	&	2.85	&	42.2	&	47.8	\\
HD 190771	&	8.8	&	33.8	&	39.4	&	\hspace{1cm}	GJ 896 A	&	1.06	&	59.1	&	23.9	\\
HD 201091	&	34.2	&	30.9	&	45	&	\hspace{1cm}	GJ 896 B	&	0.4	&	75.9	&	12.7	\\
HD 206860	&	4.55	&	25.3	&	39.4	&	\hspace{1cm}	GJ 873 (2007 July)	&	4.37	&	53.4	&	15.5	\\
HD 206860 (2007 Jul)	&	4.6	&	36.6	&	57.7	&	\hspace{1cm}	GJ 1111 (2007)	&	0.46	&	19.7	&	28.1	\\
HD 206860 (2008 Aug)	&	4.6	&	8.4	&	32.3	&	\hspace{1cm}	GJ 1111 (2008)	&	0.46	&	19.7	&	42.2	\\
HD 206860 (2010 Jul)	&	4.6	&	8.4	&	25.3	&	\hspace{1cm}	GJ 1156 (2007)	&	0.49	&	25.3	&	30.9	\\
HD 206860 (2011 Jul)	&	4.6	&	8.4	&	53.4	&	\hspace{1cm}	GJ 1156 (2009)	&	0.49	&	39.4	&	16.9	\\
HD 206860 (2013 Jul)	&	4.6	&	14.1	&	50.6	&	\hspace{1cm}	GJ 9520	&	3.4	&	42.2	&	30.9	\\
HD 106506	&	1.39	&	61.9	&	19.7	&	\hspace{1cm}	GJ 1245B (2006)	&	0.71	&	14.1	&	25.3	\\
HII 296	&	2.61	&	16.9	&	23.9	&	\hspace{1cm}	GJ 1245B (2008)	&	0.71	&	45	&	38	\\
HII 739	&	2.7	&	56.2	&	18.3	&	\hspace{1cm}	GJ 412B (2006)	&	0.78	&	25.3	&	28.1	\\
HIP 12545	&	4.83	&	59.1	&	38	&	\hspace{1cm}	GJ 412B (2009)	&	0.78	&	19.7	&	19.7	\\
HIP 76768	&	3.64	&	64.7	&	25.3	&	\hspace{1cm}	GJ 285 (2008 Jan)	&	2.77	&	33.8	&	38	\\
TYC0486-4943-1	&	3.75	&	5.6	&	42.2	&	\hspace{1cm}		&		&		&		\\
\hline
\end{tabular}
\end{minipage}
\end{table*}

\subsection{Band latitudes}
\label{seubsec:bandlat}
In this section we determine the latitude and extent of azimuthal bands of toroidal field in our sample of stars. Firstly, by examining the magnetic maps, we eliminate those stars that show no evidence of strong bands by determining the fraction of latitudes with only a single polarity in the toroidal component of the azimuthal field. From a visual inspection, stars in which this fraction is less than 0.2 do not have clear bands and we do not attempt to find a band latitude for these stars. For the remaining 67 maps, we average the field strengths over every longitude to obtain an average field strength as a function of latitude and then take the absolute value, i.e. $|\langle B_{tor,\phi}\rangle|(\lambda)$ where $\lambda$ is the latitude. We plot the latitude at which this function is maximal in Fig. \ref{fig:bandlat} as a function of rotation period for each star. Stars with maps over multiple epochs are indicated by grey lines. An indication of the band width is given by the full-width half-maximum of the peak in $|\langle B_{tor,\phi}\rangle|(\lambda)$. The size of each data point is scaled by the full-width half maximum. All points are colour coded by stellar mass. Numerical values for this plot are included in Table \ref{tab:latProperties}. It is worth highlighting that multiple azimuthal bands of opposing polarity were seen in the rapid rotators HR 1099 and AB Dor \citep{Donati2003}.

Several authors have noted that magnetic flux tends to emerge at higher latitudes on stars with shorter rotation periods as a result of a larger Coriolis force dominating over the magnetic buoyancy of the flux tubes \citep{Schuessler1992,Schuessler1996,Granzer2002}. Our results are in qualitative agreement of this statement, with the upper envelope of band latitudes in Fig. \ref{fig:bandlat} showing a decreasing trend at longer rotation periods. However, the interplay between Coriolis and buoyancy forces alone cannot explain the large range of band latitudes seen at a given rotation period. This is especially true of stars with multiple maps that show the band latitude changing significantly over the course of months/years. In some cases, the azimuthal bands even disappear and reappear between observation epochs.

\begin{figure}
	\begin{center}
	\includegraphics[trim=0.5cm 0.4cm 1cm 0.5cm, clip=true, width=\columnwidth]{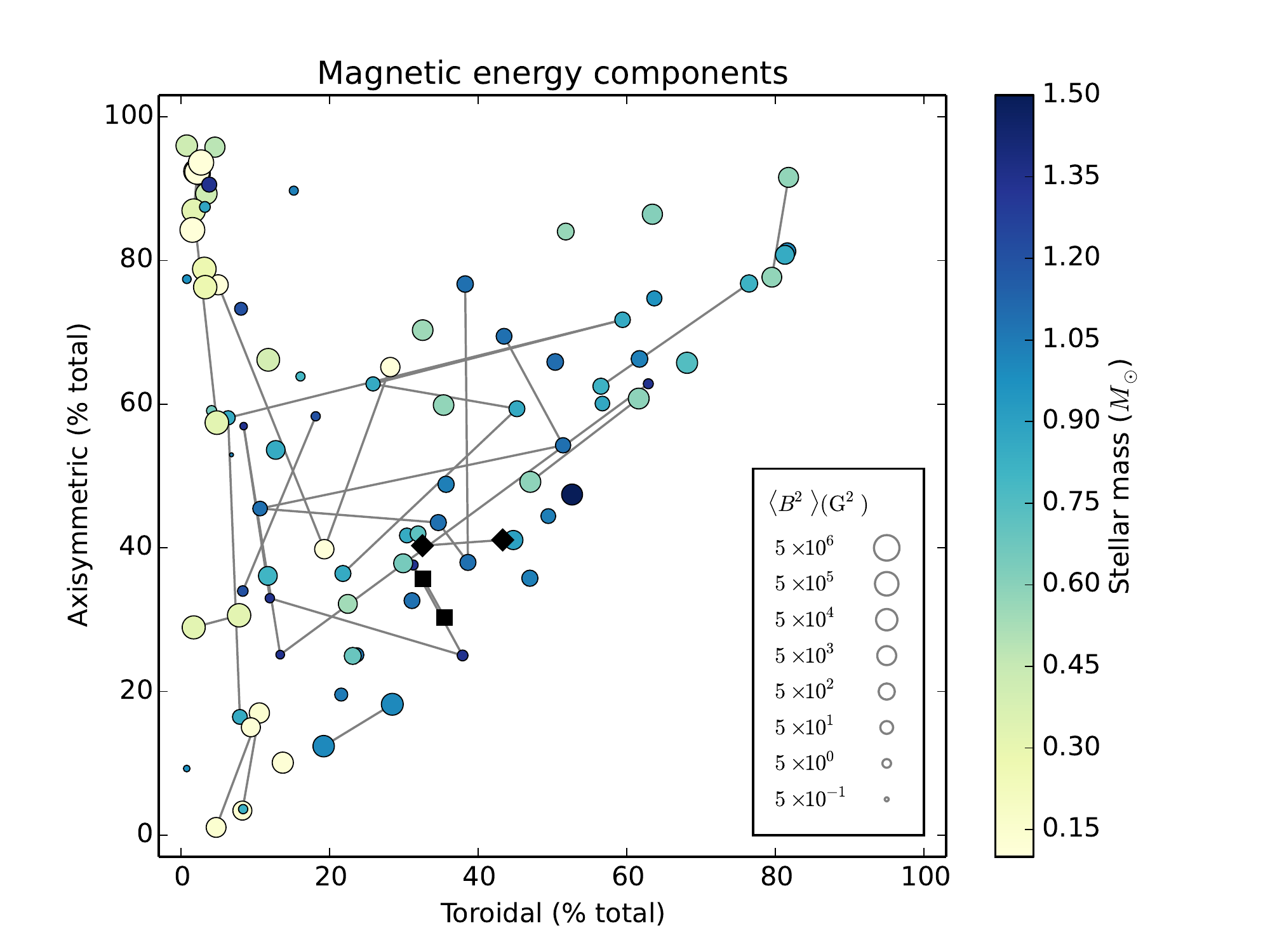}
	\end{center}
	\caption{Percentage of total magnetic energy contained in the axisymmetric component of the field against percentage of total magnetic energy contained in toroidal field. Stars observed at multiple epochs are connected by grey lines. The data points are colour coded by stellar mass and scale with the total magnetic energy density, $\langle B^2 \rangle$ (see inset). \citet{Rosen2015} reconstructed the surface field of II Peg using Stokes IV (circularly polarised light) and also Stokes IQUV (circularly and linearly polarised light). These data points are shown by black diamond and square markers respectively.}
	\label{fig:AxiTor}
\end{figure}

\subsection{Field orientation}
\label{subsec:AxisymFields}
From a visual inspection of the magnetic maps, toroidal azimuthal bands appear predominantly axisymmetrically \citep[e.g.][]{Jardine2013}. Figure \ref{fig:AxiTor} shows the fraction of magnetic energy in axisymmetric modes against the fraction contained in the toroidal component of the field. Mathematically, these correspond to the energy in the $m=0$ modes\footnote{It is worth noting that an alternative definition of axisymmetric fields, $m < \ell/2$, is sometimes used in the literature. This definition encapsulates modes that are nearly aligned with the rotation axis whereas the $m=0$ definition only accounts for modes that are exactly aligned.} and the $\gamma_{\ell,m}$ coefficients respectively. Stars with multiple epochs of observations are connected by grey lines and each point is colour coded by stellar mass. The most striking feature is the dearth of toroidal non-axisymmetric stars, in the lower right corner of the plot. Additionally, there is no trend evident with stellar mass.

The striking trend in Fig. \ref{fig:AxiTor} warrants a check for potential biases in the ZDI technique. Crosstalk between the field components, especially the radial and meridional fields is well known and has been characterised \citep{Donati1997_crosstalk}\footnote{In the tests presented by \citet{Donati1997_crosstalk}, the individual field components are considered to be independent parameters. With the new spherical harmonic implementation, as described by \citet{Donati2006}, crosstalk is considerably reduced.}. In particular, we are interested in energy from the toroidal field leaking into the poloidal field. To this end, we conducted a systematic series of tests. Specifically, we created a grid of synthetic Stokes profiles corresponding to magnetic field geometries where the axisymmetric and toroidal energy fractions ranged from zero to one. Magnetic maps were then reconstructed from these profiles and compared to the original maps that the synthetic profiles were created from. Full details of these tests can be found in Folsom et al. (in prep b). We find that there is some energy leakage from toroidal into poloidal fields (and vice versa). This effect is largest for geometries where the order, $\ell$, is equal to the degree, $m$ and smallest when there is a large difference between $\ell$ and $m$. However, the effect is not large enough to explain the dearth of points in the lower right hand corner of Fig. \ref{fig:AxiTor}.

The magnetic maps used in this study were all reconstructed using the Stokes I (unpolarised) and Stokes V (circularly polarised) profiles with one exception. \citet{Rosen2015} observed the star II Peg, during two epochs, in all four Stokes parameters (IQUV; unpolarised, linearly polarised and circularly polarised). They subsequently reconstructed magnetic maps using the more commonly used two Stokes profiles (IV) and again using all four. II Peg is plotted on Fig. \ref{fig:AxiTor} with black squares and diamonds indicating results obtained using Stokes IQUV and Stokes IV respectively. \citet{Rosen2015} found that the toroidal and axisymmetric energy fractions are similar for both reconstructions though the maps constructed with Stokes IQUV contained significantly more energy than the maps constructed from Stokes IV. The latter effect is unsurprising since one should expect more information to be reconstructed, and hence more fields, when using more data. The data points of \citet{Rosen2015} fall within the trend shown by the rest of the sample. II Peg follows this trend even though it is evolving off the main sequence whereas the rest of the sample are less evolved.

\begin{figure}
	\begin{center}
	\includegraphics[trim=1cm 1cm 1cm 0.5cm, clip=true, width=\columnwidth]{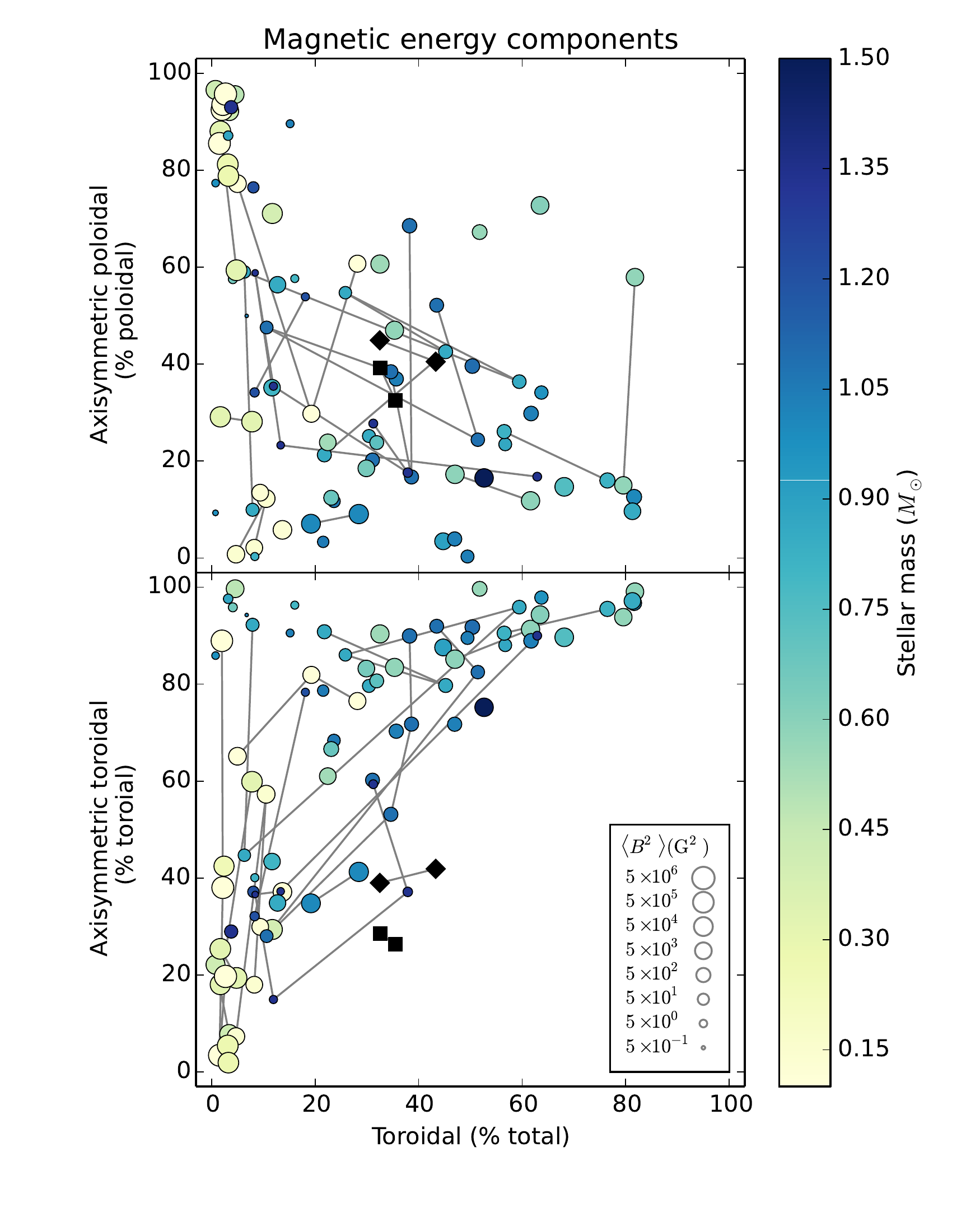}
	\end{center}
	\caption{Axisymmetric poloidal energy as a percentage of total poloidal energy (top panel) and axisymmetric toroidal energy as a percentage of total toroidal energy (bottom panel) as a function of toroidal energy. Format is the same as Fig. \ref{fig:AxiTor}}
	\label{fig:AxiTor23}
\end{figure}

Further insights may be gained by splitting the axisymmetric energy into its poloidal and toroidal components. The top panel of Fig. \ref{fig:AxiTor23} shows the fraction of poloidal energy that is axisymmetric, i.e. $E(\alpha_{\ell,m=0},\beta_{\ell,m=0})/E(\alpha_{\ell,m},\beta_{\ell,m})$, against the toroidal energy fraction. The bottom panel of Fig. \ref{fig:AxiTor23} is similar but plots the fraction of toroidal energy that is axisymmetric, i.e. $E(\gamma_{\ell,m=0})/E(\gamma_{\ell,m})$. The formatting is the same as that of Fig. \ref{fig:AxiTor}. As before, the data points of \citet{Rosen2015} for II Peg are in agreement with the rest of the sample. It is clear that the poloidal and toroidal fields behave differently. While the axisymmetric poloidal energy does not show a clear trend, a clear one is present for the axisymmetric toroidal energy. The data show that toroidal fields are generated in a preferentially axisymmetric manner. This suggests that the toroidal field generation mechanism is sensitive to the rotation axis in a way that the poloidal field is not. Noticeably, there is a cluster of M dwarfs which have dipole dominated fields in the top left hand corner of the plot (these are also present in the top left of Fig. \ref{fig:AxiTor}). \citet{Gastine2013} proposed that the strong dipolar component of these stars inhibits differential rotation and hence the generation of strong toroidal fields through the Omega effect.

\begin{figure}
	\begin{center}
	\includegraphics[trim=0.5cm 0.4cm 1cm 0.5cm, clip=true, width=\columnwidth]{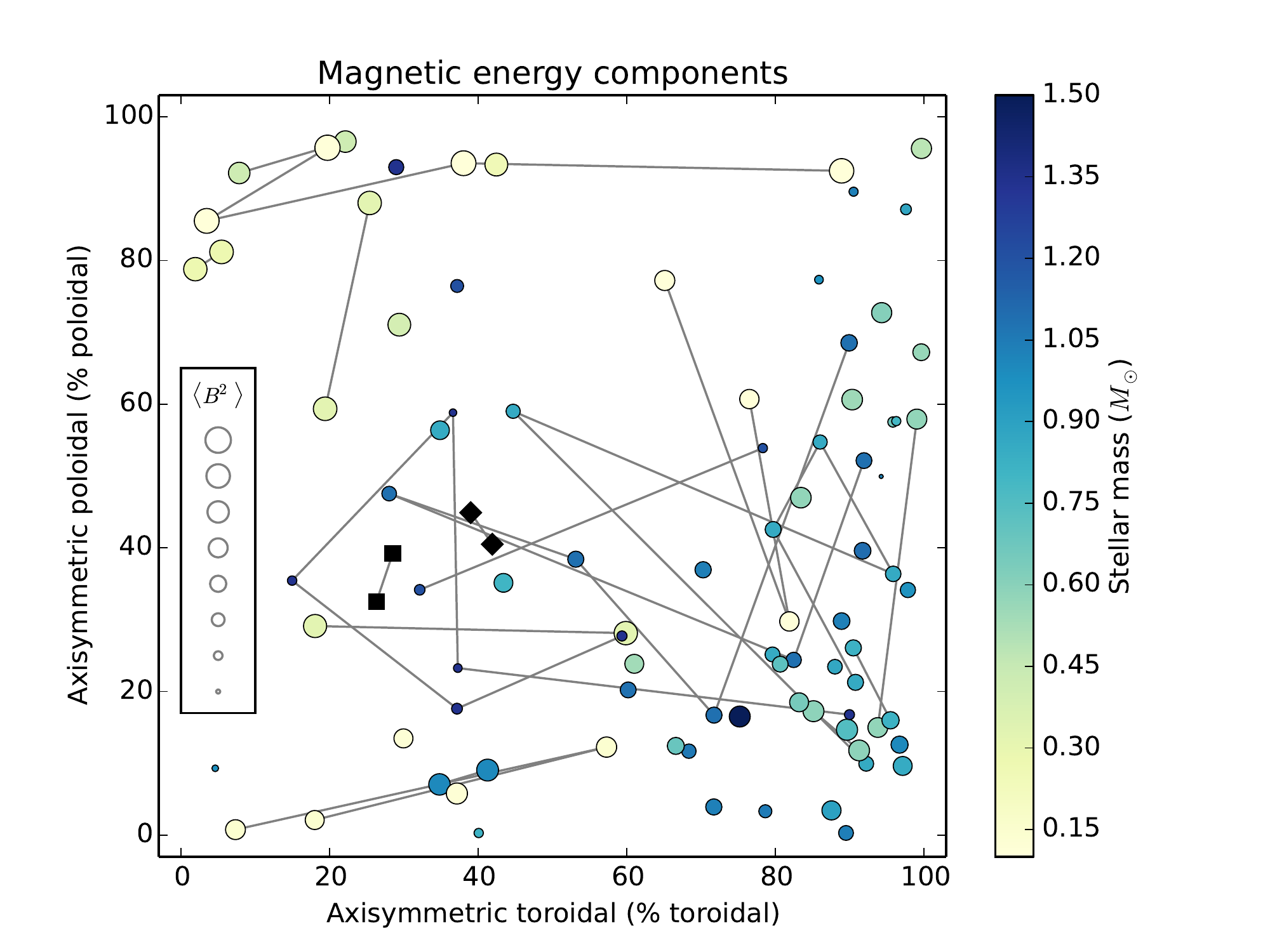}
	\end{center}
	\caption{Axisymmetric poloidal energy as a percentage of total poloidal energy against axisymmetric toroidal energy as a percentage of total toroidal energy. Format is the same as Fig. \ref{fig:AxiTor}. Similarly to Fig. \ref{fig:AxiTor}, the inset shows $\langle B^2 \rangle$ values of $5\times10^{-1} \rm{G}^2$ to $5\times10^6 \rm{G}^2$ in multiples of 10 for the smallest to biggest points. The numerical values are omitted due to lack of space.}
	\label{fig:PolAxiVsTorAxi}
\end{figure}

We can also consider how the poloidal and toroidal fields are oriented with respect to each other. Figure \ref{fig:PolAxiVsTorAxi} plots the fraction of poloidal energy that is axisymmetric against the fraction of toroidal energy that is axisymmetric, i.e. the ordinates of both panels in Fig. \ref{fig:AxiTor23}. The data points fill the entire available parameter space for this plot indicating that the poloidal and toroidal field orientations are not constrained by each other. Alternatively, their orientations may be related to each other but in a complex and non trivial way. It is particularly interesting that while the orientations of the poloidal and toroidal fields are uncorrelated, there is a relatively tight dependence of the toroidal energy density on the poloidal energy density (see Fig. \ref{fig:budget}).

\section{Conclusion}
\label{sec:Discussion}
Over the last decade, large strides have been made in characterising the strength, geometry and time evolution of cool star magnetic fields. Many stars display strong toroidal azimuthal bands, though currently there is no definitive explanation for their formation. In this paper, we explore the strength and geometry of toroidal magnetic fields by analysing 55 stars that have had their surface magnetic fields mapped with ZDI. This is the first time toroidal fields have been studied in a large sample of stars. The results presented show clear constraints for future dynamo models and can be summarised as follows.

Our sample shows that strong toroidal fields are strongly axisymmetric. Conversely, the orientation of poloidal fields are unconstrained with respect to the rotation axis. There is also evidence that the underlying dynamo type affects the relative strength of the poloidal and toroidal fields. We find that stars less massive than 0.5$\,{\rm M}_\odot$ tend to have small toroidal energy fractions while more massive stars can have substantially higher toroidal energy fractions. Additionally, the toroidal field energies of the $\rm M > 0.5\,{\rm M}_\odot$ stars have a steeper power law dependence on the poloidal field energy than those of the $\rm M < 0.5\,{\rm M}_\odot$ stars. These results are in line with the results of \citet{Donati2008a,Morin2008,Morin2010,Gregory2012} who noted changes in magnetic properties at $0.5\,{\rm M}_\odot$, roughly coinciding with the fully convective limit. This suggests some link between the break in magnetic properties with the change in internal structure. These authors also note a strong break in the amount of differential rotation at $0.5\,{\rm M}_\odot$. It is perhaps telling that the stars with the strongest toroidal fields are those that display differential rotation at their surface. On the other hand, the lowest mass stars, which do not have large toroidal energy fractions (see Fig. \ref{fig:budget}), display little to no differential rotation. This may be a hint that an Omega effect, in the form of differential rotation, is responsible for generating axisymmetric toroidal fields. It would be interesting to compare these observational results to the theoretical dynamo simulations of \citet{Brown2010} and see if their axisymmetry and poloidal/toroidal energy fractions follow the same trends.

We find that the latitude of the azimuthal bands depends, in part, on the stellar rotation period. On fast rotators, buoyant flux tubes feel a strong Coriolis force and are deflected polewards allowing flux emergence to occur at higher latitudes. Additionally, we find bands at low to mid latitudes for all rotation periods. Therefore, rotation period cannot be the only parameter determining the latitude of flux emergence. However, at present, it is unclear what further parameters may be important.

Additionally, we find that the poloidal and toroidal energies both follow the same qualitative behaviour with Rossby number. Both show saturation at low Rossby number with magnetic energies decreasing at higher Rossby number. This is also the qualitative behaviour that X-ray luminosity normalised to bolometric luminosity shows as a function of Rossby number, baring some quantitative differences such as the slope in the unsaturated regime. That the two field components behave similarly indicates that they share a common generation mechanism or that they are generated from each other. There is evidence that, in our sample, the $\rm{M}<0.5\,{\rm M}_\odot$ stars correspond to stars in the saturated regime giving credence to the claim of \cite{Wright2011} that stars in the saturated and unsaturated regimes posses different dynamo mechanisms. However, it remains to be seen whether this is an effect of the bias in our sample. Future spectropolarimetric observations of cool stars with $R_o  \lesssim  0.1$, $\rm{M}_{\star}>0.5\,{\rm M}_\odot$ will help confirm or disprove this hypothesis.

\section*{Acknowledgements}
The authors thank the anonymous referee for their constructive comments. VS acknowledges the support of an Science \& Technology Facilities Council (STFC) studentship. AAV acknowledges support from the Swiss National Science Foundation through an Ambizione Fellowship. SVJ and SBS  acknowledge research funding by the Deutsche Forschungsgemeinschaft (DFG) under grant SFB 963/1, project A16. SGG acknowledges support from the STFC via an Ernest Rutherford Fellowship [ST/J003255/1]. This study was supported by the grant ANR 2011 Blanc SIMI5-6 020 01 ``Toupies: Towards understanding the spin evolution of stars" (\url{http://ipag.osug.fr/Anr_Toupies/}). This work is based on observations obtained with ESPaDOnS at the CFHT and with NARVAL at the TBL. CFHT/ESPaDOnS are operated by the National Research Council of Canada, the Institut National des Sciences de l'Univers of the Centre National de la Recherche Scientifique (INSU/CNRS) of France and the University of Hawaii, while TBL/NARVAL are operated by INSU/CNRS. We thank the CFHT and TBL staff for their help during the observations.

\bibliographystyle{mn2e}
\bibliography{NonpotPaper}{}

\end{document}